\documentclass{iau} 
\usepackage{graphicx}

\title[Optical Spectroscopy of type1-LINERs] 
{Optical Spectroscopy of nearby type1-LINERs}

\author[Sara Cazzoli et al.]   
{S.\,Cazzoli$^1$, 
 I.\,M{\'a}rquez$^1$, J.\,Masegosa$^1$, A.\,del\,Olmo$^1$, M.\,Povi{\'c}$^{2,1}$, O. Gonz{\'a}lez-Mart{\'i}n$^3$, B.\,Balmaverde$^4$, L.\,Hern{\'a}ndez-Garc{\'i}a$^5$ and S. Garc{\'i}a-Burillo$^6$}

\affiliation{
$^{1}$ IAA - Instituto de Astrof{\'i}sica de Andaluc{\'i}a (CSIC), Apdo. 3004, 18080, Granada, Spain \\ email: {\tt sara@iaa.es} \\[\affilskip]
$^{2}$ ESSTI/EORC - Ethiopian Space Science and Technology Institute, Entoto Observatory and Research Center,   P.O. Box 33679, Addis Ababa, Ethiopia\\ 
$^{3}$ IRyA - Instituto de Radioastronom{\'i}a y Astrof{\'i}sica, 3-72 Xangari, 8701, Morelia, Mexico\\ 
$^{4}$ INAF - Osservatorio Astronomico di Brera, via Brera 28, I-20121 Milano, Italy\\
$^{5}$ Universidad de Valpara{\'i}so, Gran Bretana 1111, Playa Ancha, Valpara{\'i}so, Chile\\ 
$^{6}$ OAN - Observatorio Astron{\'o}mico Nacional, Alfonso XII, 3, 28014, Madrid, Spain 
}

\pubyear{2019}
\volume{356}  
\jname{Nuclear activity in galaxies across cosmic time}
\editors{M. Povic, J. Masegosa, H. Netzer, P. Marziani, \\ P. Shastri, S. B. Tessema $\&$ S. H. Negu,  eds.}
\begin{document}

\maketitle

\begin{abstract}
We present the highlights from our recent study of 22 local (z\,$<$\,0.025) type-1 LINERs from the Palomar Survey, on the basis of optical long-slit spectroscopic observations taken with TWIN/CAHA, ALFOSC/NOT and HST/STIS. Our goals were threefold: (a)\,explore the AGN-nature of these LINERs by studying the broad (BLR-originated) H$\alpha$$\lambda$6563 component; (b)\,derive a reliable interpretation for the multiple narrow components of emission lines by studying their kinematics and ionisation mechanism (via standard BPTs); (c)\, probe the neutral gas in the nuclei of these LINERs for the first time. Hence, kinematics and fluxes of a set of emission lines, from H$\beta$$\lambda$4861 to [SII]$\lambda$$\lambda$6716,6731, and the NaD$\lambda$$\lambda$5890,5896 doublet in absorption have been modelled and measured, after the subtraction of the underlying light from the stellar component.
\end{abstract}

\firstsection 

\section{Introduction}
\noindent  Low ionisation nuclear emission-line regions (LINERs) are a class of low-luminosity AGNs  showing strong low-ionisation and faint high-ionisation emission lines (\cite[Heckman 1980]{Heckman1980}).  LINERs are  interesting objects  since they are the most numerous local AGN population bridging the gap between normal and active galaxies (\cite[Ho\,2008]{Ho2008}). \\ 
Over the past 20 years, the ionising source in LINERs has been studied through a multi-wavelength approach via different tracers (\cite[Ho\,2008]{Ho2008}). Nevertheless, a long standing issue  is the origin and excitation mechanism of the ionised gas studied via optical emission lines.  In addition to the AGN scenario, two more alternatives, such as pAGBs stars and shocks, have been proposed to  explain the  optical properties of LINERs (e.g.\,\cite[Singh\,et\,al.\,2013]{Singh2013}). \\ 
In LINERs, outflows  are  common as suggested by their H$\alpha$ nuclear morphology (\cite[Masegosa et al. 2011]{Masegosa2011}). To open a new window to explore the AGN-nature  and the excitation mechanism in LINERs,  we  propose to infer the role of outflows (identified as relatively broad component) in the broadening of emission lines. This broadening effect may limit the spectroscopic classification, as the contribution of outflows may overcome the determination of an eventually  faint and  broad  H$\alpha$ component from the BLR.  \\ 
Ionised gas outflows are observed in starbursts and AGNs via long slit (e.g.\,\cite[Harrison\,et\,al. 2012]{Harrison2012}) and integral field spectroscopy  (e.g.\,\cite[Maiolino et\,al.\,2017]{Maiolino2017}) of emission lines. The neutral gas in outflows have been studied in detail only in starbursts and luminous and ultra-luminous infrared galaxies (U/LIRGs) via the NaD absorption (e.g. \cite[Cazzoli\,et\,al.\,2016]{Cazzoli2016}).\\
\noindent In \cite[Cazzoli et al. 2018]{Cazzoli2018}, hereafter C18, our goals were to investigate the AGN nature of type-1 LINERs and to characterize all the components by studying their kinematics and ionisation mechanisms. We also aimed to probe and study  the neutral gas properties. \\
 For type-2 LINERs, see the contribution by L.\,Hermosa-Mu\~{n}oz in this volume.
 
\vspace{-0.15cm}
\section{Sample, Data and Methods} 
\label{sampledatamethods}
\noindent The sample contains nearby (\textit{z}\,$<$\,0.025) 22  type-1 LINERs (L1) selected from the Palomar Survey (see  \cite[Ho\,et\,al.\,1997]{Ho1997}).  Spectroscopic data were gathered with the TWIN Spectrograph mounted on the 3.5m telescope of the Calar Alto Observatory (CAHA) and with ALFOSC attached to the 2.6m North Optical Telescope (NOT). We also analyzed archival spectra (red bandpass)  for 12 LINERs (see \cite[Balmaverde\,et\,al.\,2014]{Balmaverde2014} for details) obtained with the Space Telescope Imaging Spectrograph (STIS) on board the Hubble Space Telescope (\textit{HST}). The data analysis is organised in three main steps:\\
\noindent \underline{\textit{Stellar Subtraction}} \\ 
We applied the penalized PiXel fitting  (\textsc{pPXF}; \cite[Cappellari 2017]{Cappellari2017}) and  the \textsc{starlight} methods (\cite[Cid\,Fernandes\,et\,al.\,2009]{CidFernandes2009}) for  modeling the stellar continuum. The stellar model is then subtracted to the observed to one obtain a interstellar medium spectrum.\\ 
\noindent \underline{\textit{Emission Lines}}\\
The fit was performed simultaneously for emission lines from [OI]$\lambda$$\lambda$6300,6363 to [SII], with  single or multiple Gaussian kinematic components (up two for forbidden lines and narrow H$\alpha$).  For the modeling of the H$\alpha$-[NII]$\lambda$$\lambda$6548,6584 blend, we tested three distinct models. Specifically, we considered either [SII] or [OI] (\texttt{S}- and  \texttt{O}- models) or both (\lq mixed\rq \, \texttt{M}-model) as  reference for tying central wavelengths and line widths (see example in Fig.\,1).  The latter  model takes into account possible stratification density in the narrow line region. Then, a broad H$\alpha$  component  is added if needed to reduce significantly the residual. Finally, the best fitting (i.e. model, components, velocity shift and line widths) has been constraint to be the same for [OIII]$\lambda$$\lambda$4959,5007 and H$\beta$ lines. Intensity ratios for [NII], [OI]  and [OIII]  lines were imposed following  \cite[Osterbrock\,2006]{Osterbrock2006}.  \\
\noindent \underline{\textit{Absorption Lines}}\\
The NaD absorption doublet (8 detections) was modelled with one (i.e. two Gaussian profiles) or two components as in \cite[Cazzoli\,et\,al.\,2014]{Cazzoli2014}. The ratio of the  equivalent widths of the two lines was allowed to vary from 1 to 2 (i.e. optically thick/thin limits, \cite[Spitzer\,1978]{Spitzer1978}).\\

\vspace{-0.6cm}
\section{Main Results}
\label{mainresults}
\noindent NGC\,4203 represents an extreme case as three line components are not sufficient to reproduce well the H$\alpha$  profile, therefore, we excluded this L1 from the analysis.\\
\noindent  For ground-based data, the \texttt{S}- and  \texttt{O}- models  reproduce well  the line profiles in six of the cases each, while a larger fraction of cases (i.e. 9/21) require \texttt{M}-models for a satisfactory fit. Of the four  possible combinations of the three components, as single narrow Gaussian per forbidden line is adequate in  6 out of 21 cases.  A broad H$\alpha$ component is required in the remaining four cases. In most cases (15/21), two Gaussians per forbidden line are required for a satisfactory modelling. Among these 15 cases, only in three cases a  broad H$\alpha$ component is required to  reproduce well the observed profiles.  \\ 
Velocities of the narrow components  are close to rest frame varying within $\pm$\,110 km\,s$^{-1}$. The average velocity dispersion value for the narrow components is $\sigma$\,=\,157\,km\,s$^{-1}$. For the second components the velocity range is large, from -350 km\,s$^{-1}$ to 100 km\,s$^{-1}$. The velocity dispersion varies  between 150 and 800 km\,s$^{-1}$ being generally broader (on average $\sigma$\,=\,429\,km\,s$^{-1}$) than for narrow components. A broad  H$\alpha$ component is required only in 7 out of the 21 LINERs, with FWHMs  from 1277\,km\,s$^{-1}$  to 3158\,km\,s$^{-1}$. \\ %
\noindent For none of the 11 \textit{HST}/STIS spectra, the adopted best fit is obtained using the   \texttt{O}-model, finding a slightly large prevalence of  best fits  with \texttt{M}-models. In four cases, one Gaussian per forbidden line and narrow H$\alpha$ is adequate. In the remaining cases, two Gaussians are required for a good fit. \\
Narrow components have velocities between -100 and 200  km\,s$^{-1}$; the velocity dispersions vary  between  120 and 270 km\,s$^{-1}$ (176\,km\,s$^{-1}$, on average). Similarly,  the velocities of second component range from -200 to 150  km\,s$^{-1}$. These second components are however broader, with   velocity dispersion values  between 300 and 750 km\,s$^{-1}$ (433 km\,s$^{-1}$, on average). The broad component in  is ubiquitous. The  FWHM of the broad H$\alpha$ components in \textit{HST} spectra  range from 2152\,km\,s$^{-1}$ to 7359\,km\,s$^{-1}$ (3270\,km\,s$^{-1}$, on average).\\
\noindent For the NaD absorption, in 7 out of 8 targets, a single kinematic component gives a good fit. Velocities of the neutral gas narrow components vary between  -165 and 165 km\,s$^{-1}$;  velocity dispersions values are in the range 104-335 km\,s$^{-1}$ (220 km\,s$^{-1}$, on average). 

\vspace{-0.15cm}
\section{Discussion}
\begin{figure}
\begin{center}
 \includegraphics[width=.92\textwidth]{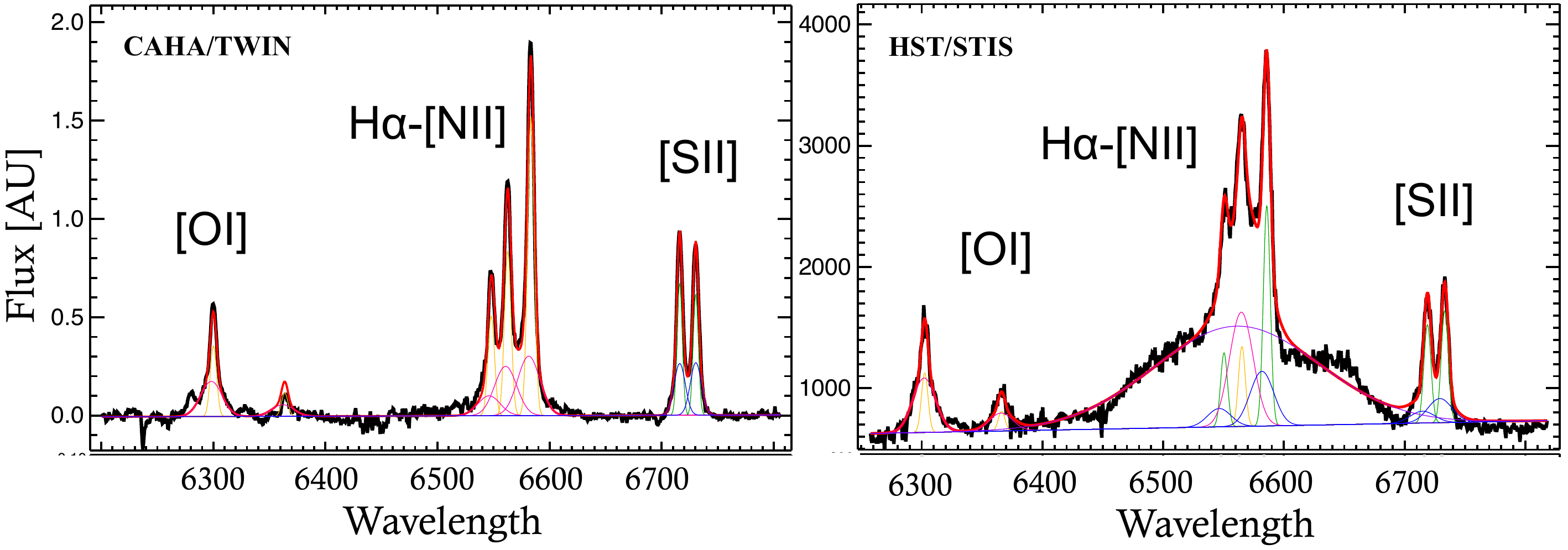} 
 \vspace*{-0.2 cm}
 \caption{Gaussian fit to the ground- (left) and space- (right) based spectra for NGC4450 in the  H$\alpha$  region. We marked with different colours the components  required to model the emission lines.  The red curve shows the total contribution from the fit. }
   \label{fig1}
\end{center}
 \vspace*{.15 cm}
\begin{center}
 \includegraphics[width=.95\textwidth]{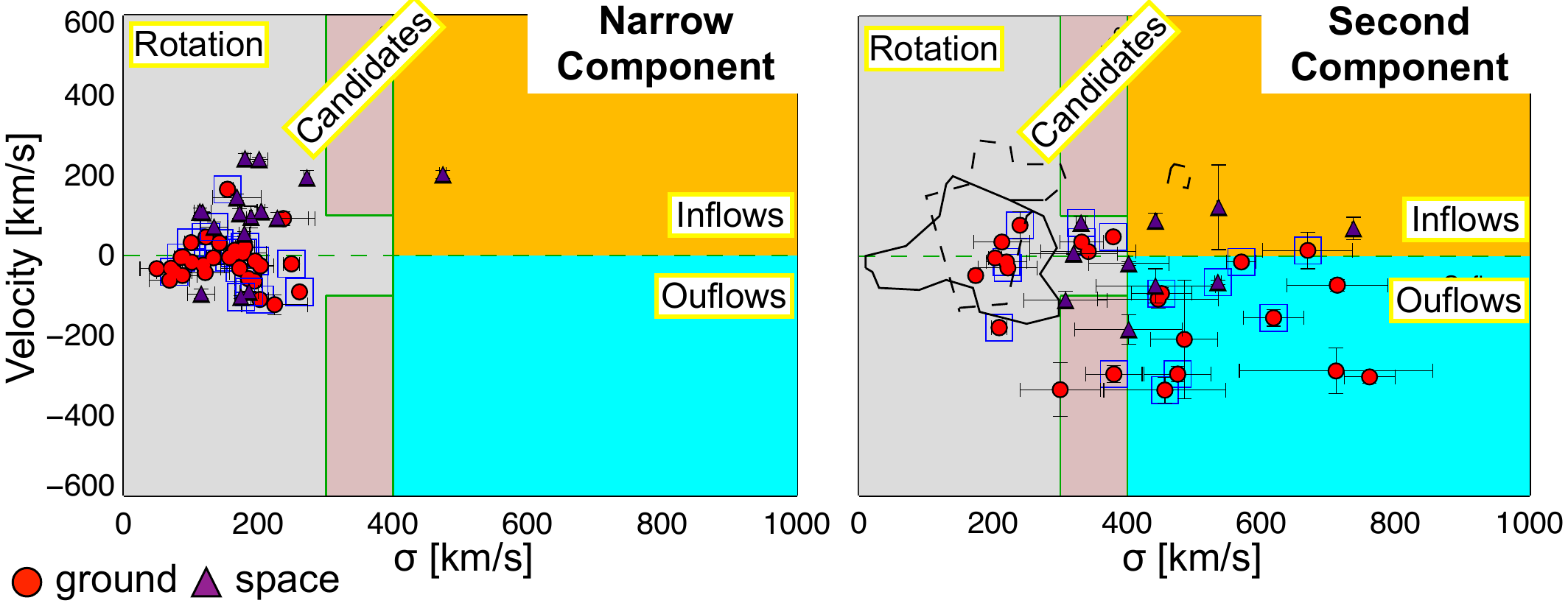} 
 \vspace*{-0.1 cm}
\caption{Observed velocity dispersion\,-\,velocity plane for narrow (left) and second (right) components. Red circles an purple triangles mark the measurements from ground- and space-based data, respectively. An additional blue box marks those LINERs for which the fitting of the H$\alpha$ profile is less reliable (see C18).  In the right panel, we report  the measurements of the narrow component with contours (continuos and dashed lines are for ground- and space-based data, respectively). The coloured areas   indicate different classifications (labelled on the panels). } 
   \label{fig2}
\end{center}
 \vspace*{.15 cm}
\begin{center}
 \includegraphics[width=.975\textwidth]{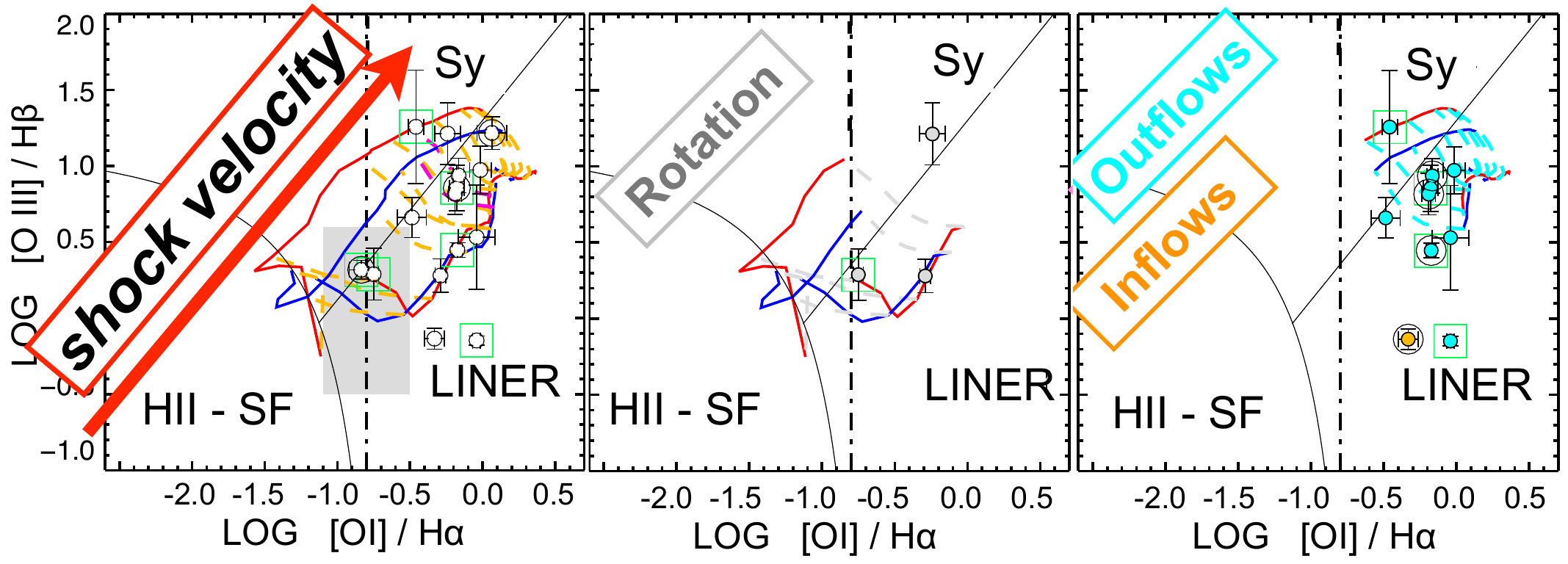} 
 \vspace*{-0.1 cm}
\caption{\textit{Left:} optical standard [OI] BPT diagram (\cite[Baldwin\,et\,al.\,1981]{Baldwin1981}) for the second component obtained from our ground-based spectroscopy (circles). In all the panels an additional circle marks those LINERs for which  a broad component is required to reproduce the H$\alpha$ profile. Light-green square are as in Fig.\,2. Black lines represent the dividing curves between HII regions, Seyferts, and LINERs from \cite[Kewley\,et\,al.\,2006]{Kewley2006} and \cite[Kauffmann\,et\,al.\,2003]{Kauffmann2003}, and weak-[OI] and strong-[OI] LINERs from \cite[Filippenko\,$\&$\,Terlevich\,1992]{Filippenko1992}. Gray boxes show the predictions of photoionisation models by pAGB stars by  \cite[Binette\,et\,al\,1994]{Binette1994}. The predictions of  shock+precursor -ionisation models from \cite[Groves et al.2004]{Groves2004}  with  n$_{\rm el}$\,=\,100\,cm$^{-3}$ (blue) and n$_{\rm el}$\,=\,1000\,cm$^{-3}$ (red) are overlaid. Iso-velocities are marked with yellow dashed-lines. \textit{Center} and  \textit{Right} panels are the same but considering considering different shock-velocities (colour coded according to Fig.\,\ref{fig2}).} 
   \label{fig3}
\end{center}
\end{figure}

\noindent \underline{\textit{Probing the BLR in L1}} \\ 
The analysis of Palomar spectra by \cite[Ho\,et\,al.\,(1997)]{Ho1997} indicated that all the LINERs in our selected sample show a broad H$\alpha$ component resulting in their classification as L1. Nevertheless, for ground-based data our detection rate for the  broad  component is only 33\,$\%$, questioning the classification as L1 by \cite[Ho\,et\,al.\,(1997)]{Ho1997}. For space-based data the broad  component is ubiquitous in agreement with \cite[Balmaverde\,et\,al.\,2014]{Balmaverde2014}. By comparing our strategy and measurements with those by  previous works (Sect.\,5.1 in C18), we  conclude that the detectability of the BLR-component is sensitive to  the starlight decontamination and the choice of the template for the H$\alpha$-[NII] blend. Moreover, a single Gaussian fit for the forbidden lines is an oversimplification in many cases.  \\
\noindent \underline{\textit{Kinematic classification of the components}} \\ 
The distribution of the velocity for narrow and second components as a function of their velocity dispersion is presented in Fig.\,\ref{fig2}. We identified  four areas corresponding to different kinematical explanations: rotation, candidate for non-rotational motions and non-rotational-motions (with  broad blue/redshifted lines produced by outflows/inflows).  For both ground- and space-based data, the kinematics of the narrow component can be explained with rotation in all cases (Fig.\,2\,left) whereas that of the second components encompass all possibilities (Fig.\,2, right). From our ground-based data, we identified 6 out of 15 (40\,$\%$) cases that may be interpreted as outflows. Outflow-components have velocities varying from -15\,km\,s$^{-1}$ to -340\,km\,s$^{-1}$, and velocity dispersions in the range of 450-770\,km\,s$^{-1}$ (on average, 575\,km\,s$^{-1}$).   We did not interpreted as outflows any case in \textit{HST}/STIS data. These results  partially disagree with studies of the H$\alpha$ morphology in LINERs which indicate that outflows are frequent in LINERs. A possible explanation is that the extended nature of outflows is not fully captured by the \textit{HST}/STIS spectra.\\
\noindent \underline{\textit{Ionisation mechanisms from standard \lq BPT (Baldwin,\,Phillips\,\&\,Terlevich)-diagrams\rq}} \\ 
\noindent The line ratios for the narrow component are generally consistent with those observed in AGNs (either  Seyfert  or LINERs), excluding  the star-formation or pAGBs as the dominant ionization mechanism (see Fig.\,9 in C18). For the second component, we reproduced the observed line ratios with the shock-models by \cite[Groves et al.\,2004]{Groves2004}. We combined the data points and models for the [O\,I]/H$\alpha$ BPT diagram (\cite[Baldwin\,et\,al.\,1981]{Baldwin1981}),  the most reliable for studying shocks (\cite[Allen\,et\,al.\,2008]{Allen2008}) shown in  Fig.\,\ref{fig3}, with the kinematical classification shown in Fig.\,\ref{fig2}. 
Models at low velocities ($<$\,300\,km\,s$^{-1}$) indicate the presence of mild-shocks associated to perturbations of rotation (Fig.\,3 center). At higher velocities ($>$\,400\,km\,s$^{-1}$) shocks are produced by non-rotational motions (Fig.\,3\,right).\\
\noindent \underline{\textit{A lack of neutral outflows?}} \\ 
According to the adopted kinematic classification all the neutral gas kinematic components (except one) could be interpreted as rotation. The possible explanation of the lack of neutral gas non-rotational motions, such as outflows, is twofold. First, the neutral component in outflows is possibly less significant  in AGNs than in starbursts galaxies and U/LIRGs. Secondly, such a null detection rate might be a consequence of the conservative limits we assumed, as ionised and neutral gas correspond to different phases of the outflows, and hence these may have a different kinematics (e.g. \cite[Cazzoli et al. 2016]{Cazzoli2016}). 

\vspace{-0.15cm}
\section{Conclusions}
\noindent $\bullet$ \underline{\textit{The AGN nature of L1.}} The  detection of the BLR-component is sensitive to  the starlight subtraction,  the template for the H$\alpha$-[NII] blend and the assumption of a single Gaussian fit (often an oversimplification); NLR stratification might be often present in L1.\\ 
\noindent $\bullet$ \underline{\textit{Kinematics of emission lines and their classification.}}  The kinematics of the narrow component can be explained with rotation in all cases whereas that of the second components encompass all possibilities. From our ground-based data, the detection rate of outflows is 40\,$\%$. We did not interpreted as outflows any case in \textit{HST}/STIS data. \\ \noindent $\bullet$ \underline{\textit{Ionisation mechanisms.}}  Our results favor the  AGN photoionisation as the dominant mechanism of ionisation for the narrow component.  Shocks models (at the observed velocities) are able to reproduce the observed line ratios of the second component.\\
\noindent $\bullet$ \underline{\textit{Neutral gas in L1.}} The neutral gas is found to be in rotation. Neutral gas outflows are possibly less significant  in AGNs or the limits we assumed are too conservative.

\vspace{-0.15cm}
\section*{Acknowledgements}
\noindent We thank the financial support by the Spanish MCIU and MEC, grants SEV-2017-0709 and AYA 2016-76682-C3. SC thanks the IAU for the travel grant.

\end{document}